\documentclass[
    ,final            
  ]
  {aipproc}

\layoutstyle{8x11single}

\usepackage{amsmath}
\usepackage{epsfig}
\begin{document}

\title{On a Missed Mechanism of Dielectron Production in
Nucleon-Nucleon Collisions}

\classification{25.75-q; 25.75.Dw; 13.75.Cs; 14.20.Pt; 13.30Ce}
\keywords      {DLS puzzle; dilepton spectroscopy; mechanisms of dilepton production; dibaryon}

\author{Anatoliy S.Khrykin}{
  address={Dzhelepov Laboratory of Nuclear Problems, JINR, 141980 Dubna, Moscow Region, Russia}
}

\begin{abstract}
We examine a new mechanism of $e^+e^-$ pair production in $NN$ collisions associated with the $NN$-decoupled
dibaryon $d^\star_1$(1956) formation in the process $NN \to \gamma^\star d^\star_1$, where $\gamma^\star$ is
the virtual photon which converts into a $e^+e^-$ pair. It is shown that a substantial excess of
dielectron yields
from $Ca+Ca$ and $C+C$ collisions at 1 GeV/A in the dielectron mass spectra  in the region from 0.2 to 0.5 GeV/$c^2$ measured by
the \textit{DLS Collaboration} as compared with calculated ones can be attributed to the contribution of this mechanism.
A simple means for verification of the existence of such a mechanism is proposed.
\end{abstract}

\maketitle


\section{Introduction}
Experimental results on dielectron ($e^+e^-$ pair) production in collisions of $Ca+Ca$, and $C+C$
\cite{DLSAAex} at 1.04 GeV/A and $p+d$, and $p+p$ at a number of energies from 1 to 4.88 GeV/A \cite{DLSpppdex}
reported by the \textit{DLS
Collaboration} more than 10 years ago have demonstrated that theoretical models at the time (see,e.g. Refs. \cite{Ernst,BRCas98,FFKMppeeX}
failed to give a satisfactory account of the measured $e^+e^-$ pair invariant mass spectra.
In particular, such spectra from $Ca+Ca$, and $C+C$ collisions substantially exceed those predicted by the model
calculations in the dielectron mass $M$ region $0.2 < M < 0.5$ GeV/$c^2$
and reference therein).
This \textit{DLS} finding called the "DLS puzzle" was recently confirmed by the \textit{HADES Collaboration}
\cite{HADES08} and now it is regarded as firmly established.

Since the shortage of dielectrons in calculated spectra became apparent,
several theoretical groups have attempted to understand its origin. The emphasis mainly concentrated on
the development of suitable models describing the conventional elementary processes
of $e^+e^-$ pair production \cite{Kaptbremst}. These processes are believed to be associated with the production in $NN$ collisions of neutral mesons (m), such as $\pi^0$, $\eta$,
$\omega$,..., and baryon resonances (R), such as $\Delta(1232)$, $N$(1520),..., which can undergo
the Dalitz decay $m(R)\to \gamma e^+e^-(N e^+e^-)$, or the vector mesons $\rho^0$, and $\omega$ which can directly decay into
into $e^+ e^-$ pairs, and the $NN$ virtual (timelike) photon bremsstrahlung $NN \to NN\gamma^\star\to NN e^+e^-$.
Nevertheless, despite considerable efforts the nature of the discrepancy between the calculated and measured
spectra noted above  is still obscure.

The shortage of dielectrons in calculated spectra could
indicate the existance of an additional source of $e^+e^-$ pairs that
was not included in the model calculations. One of such possible sources is due to the $NN$-decoupled
dibaryon resonance $d^\star_1$(1956).
This resonance \cite{PRC64} can be produced in the radiative process $NN \to
\gamma d^{\ast }_{1}$ and decays in that
$d^{\ast }_{1} \to NN \gamma$. It has been found that it manifests itself in a series of
experimentally studied photon production processes induced by nucleon-nucleon and nucleon-nucleus
collisions at intermediate energies \cite{NPA721,Menu07}.
However, we know little about the physical nature of this resonance
and do not yet know its quantum numbers. We only know that its spin $J$ and parity $P$ can be
either from a set $J^P=1^+,3^+$, etc. if its isospin $I=1$ or any if $I=2$.

At the same time, if the resonance $d^\ast_1$ really exists then together with
its formation (decay) channel with a real photon
the same channels with a virtual, massive photon should also take place.
Conversion of such photons into $e^+ e^-$ pairs
would give rise to a new source of dielectrons.
However the virtual photons from the $d^\star_1$ decay should have relatively small masses which do not exceed
the kinematical limit $M_{max}=M_R-M_{NN} \sim 0.08$ GeV$/c^2$, where $M_R$ and $M_{NN}$ are the mass
of the resonance and the sum of masses of nucleons from this decay, respectively.
Therefore this process cannot have to do with the shortage of dielectrons in question.
In this paper we examine the $e^+e^-$ pair production mechanism associated with the $d^\star_1$ formation in $NN$ collisions
due to virtual photon emission and show
that its contribution to the invariant mass spectra of the dielectron from $p+p$, $Ca+Ca$, and $C+C$ collisions
at 1.0 GeV/A  can supply the observed shortages of $e^+ e^-$ pairs in these spectra
 in the region $0.2 < M < 0.5$ GeV/$c^2$.

\section{Kinematics and Matrix element}
The process to be considered is
\begin{equation}
N(p_1)+N(p_2)\to \gamma^\star(k)+ d^\star_1 \to e^+(p_3) + e^-(p_4) + d^\star_1(p_5),
\end{equation}
with $p_1+p_2=p_3+p_4+p_5$,
where $p_1$ and $p_2$ are the four-momenta of the colliding nucleons,
$p_3$, $p_4$, $p_5$ are the four-momenta of the electron, the positron and
the dibaryon, respectively, and $k=p_3+p_4$ is the four-momentum of the virtual photon.
The kinematical region for its mass $M$ is $2m_e\le M
\le M_{max}=\sqrt{s} - M_R$, where $m_e$ is the electron mass and $\sqrt{s}$ is
the total CMS energy of colliding nucleons. This mass region overlaps with that where the shortage of dielectrons of interest was observed. For example,
in the case of $pp$ collisions at a kinetic energy of the incident protons $T_k= 1.04$ GeV the upper
limit for the dielectron mass is $M_{max}\simeq 0.38$ GeV.

The matrix element ${\cal{M}}$ for the process (1) can be written as
\begin{align}
{\cal{M}}=\frac{e^2}{k^2}j^\mu J_\mu,
\end{align}
where $J_\mu=(J_0,\vec{J})=<d^\star_1|\hat{J_\mu}|NN>$ and $j_\mu=(j_0,\vec{j})=<e^+ e^-|\hat{j_\mu}|0>$ are the hadronic and leptonic electromagnetic transition currents,
 respectively, and $\hat{J_\mu}$ and $\hat{j_\mu}$ are the operators of these currents.
Note that the same current $J_\mu$ also describes the process $N+N\to \gamma+ d^\star_1$,
where $\gamma$ is a real photon. In this case, however, the space part $\vec{J}$ of this
current is always perpendicular to the momentum $\vec{k}$ of the photon.
The space part $\vec{J}$ of the current $J_\mu$ for the process (1)
can, in general, have both a transverse $\vec{J_T}$  and a longitudinal $\vec{J_L}$ ($\vec{J_L}\|\vec{k}$) component, so that $\vec{J}=\vec{J_L}+\vec{J_T}$.

The square of the matrix element (2) can be represented in terms of a transverse and a longitudinal components of  $\vec{J}$ as\cite{NPA574}:
\begin{align}
|{\cal{M}}|^2&= \frac{1}{4}\frac{e^4}{M^4}\frac{1}{2{m_e}^2} (M^2|\vec{J_T}|^2-|\vec{J_T}\vec{q_T}|^2+\frac{M^2}{k_0^2}
(1-\frac{M^2}{k_0^2}|\vec{q_L}\vec{J_L}|^2
-2\frac{M^2}{k_0^2}|\vec{q_L}\vec{J_L}||\vec{q_T}\vec{J_T}|,
\end{align}
where $\vec{q_L}$ and $\vec{q_T}$ are the longitudinal and transverse components of the momentum $\vec{q}=\vec{p_3}-\vec{p_4}$, respectively.

The quantities $\vec{J_T}$ and $\vec{J_L}$ in Eq. (3) depend on the quantum numbers of the $|d^\star_1>$ and its electromagnetic structure. We assume in this work that the $d^\ast_1$  has spin zero and isospin two.
The general structure of the electromagnetic transition from an initial $|pp>$ state to a final pointlike $|d^\star_1>$ state with such quantum numbers has been investigated in the work  \cite{Scholton} and found to be of the magnetic type.
The square of the matrix element for the magnetic dipole ($M1$) transition $p+p\to \gamma + d^\star_1$ derived in the \cite{Scholton} is given by
\begin{equation}
 |{\cal{M}}(pp \to \gamma d^\star_1)|^2=e^2|\vec{J_T}|^2={C}e^2\cdot(p_1 \cdot
 k)(p_2\cdot k),
\end{equation}
 where $p_1$ and $p_2$  are the
four-momenta of the colliding nucleons, and $C$ is a certain normalization constant.

The fundamental property of an electromagnetic current for the magnetic type transition with a real or virtual photon is that its components
 satisfy the conditions $J_0=0$ and $\vec{J}\bot\vec{k}$.
The square of the matrix element (3) for such a transition between an initial state $|NN>$ and a final state $|d^\star_1>$ takes the form
\begin{align}
|{\cal{M}}|^2&=\frac{1}{4}\frac{e^2}{M^2}\frac{1}{2{m_e}^2} \left(1-\frac{1}{2}\frac{|\vec{q_T}|^2}{M^2}\right) e^2|\vec{J_T}|^2
\end{align}
\section{spectra calculations and comparison with DLS data}
 The differential cross section $d\sigma /dM$ for the production of a dielectron
 of invariant mass $M$ in the process (1) is
given by
\begin{equation}
\frac{d\sigma}{d M}=\frac{2 \pi^4}{4f}\int |{\cal M}|^2|F(M^2)|^2
\prod_{i=3}^{5}\frac{d^3\vec{p_i}}{2E_i
2\pi^3}\delta^4(p_1+p_2-\sum_{i=3}^{5}p_i)\delta(M-M(p_3,p_4)),
\end{equation}
 where $f=\sqrt{(p_1p_2)^2-m_1^2m_2^2}$, $m_1$ and $m_2$ are the masses of the colliding nucleons and $F(M^2)$ is the form factor for the $|NN> \to |d^\star_1>$ transition.
 In our calculation we used the form factor predicted by the vector dominance model:
\begin{equation}
F(M^2)=\frac{{m_\rho}^4+{m_\rho}^2{\Gamma_\rho}^2}{({m_\rho}^2-M^2)^2+{m_\rho}^2{\Gamma_\rho}^2},
\end{equation}
where $m_\rho$ and $\Gamma_\rho$ are the mass and width of the $\rho$ meson, respectively.

The integral (6) was calculated by the Monte Carlo method. The phase space region of the integration was limited by the
{\it DLS} apparatus acceptance that was given by the corresponding filter provided by the {\it DLS collaboration}. In calculations the finite mass resolution of the {\it DLS} spectrometer $\sigma_M/M=0.1$ was taken into account.
In the case of $C+C$ collisions the calculation was done within the quasi-free $NN$ collision approximation. To allow for the Fermi motion of nucleons in a
$^{12}C$ nucleus the nucleon momentum distribution  obtained in the work \cite{Cmom} was used.
\begin{ltxfigure}[htb]
\begin{minipage}[c]{80mm}
  \includegraphics[height=.30\textheight]{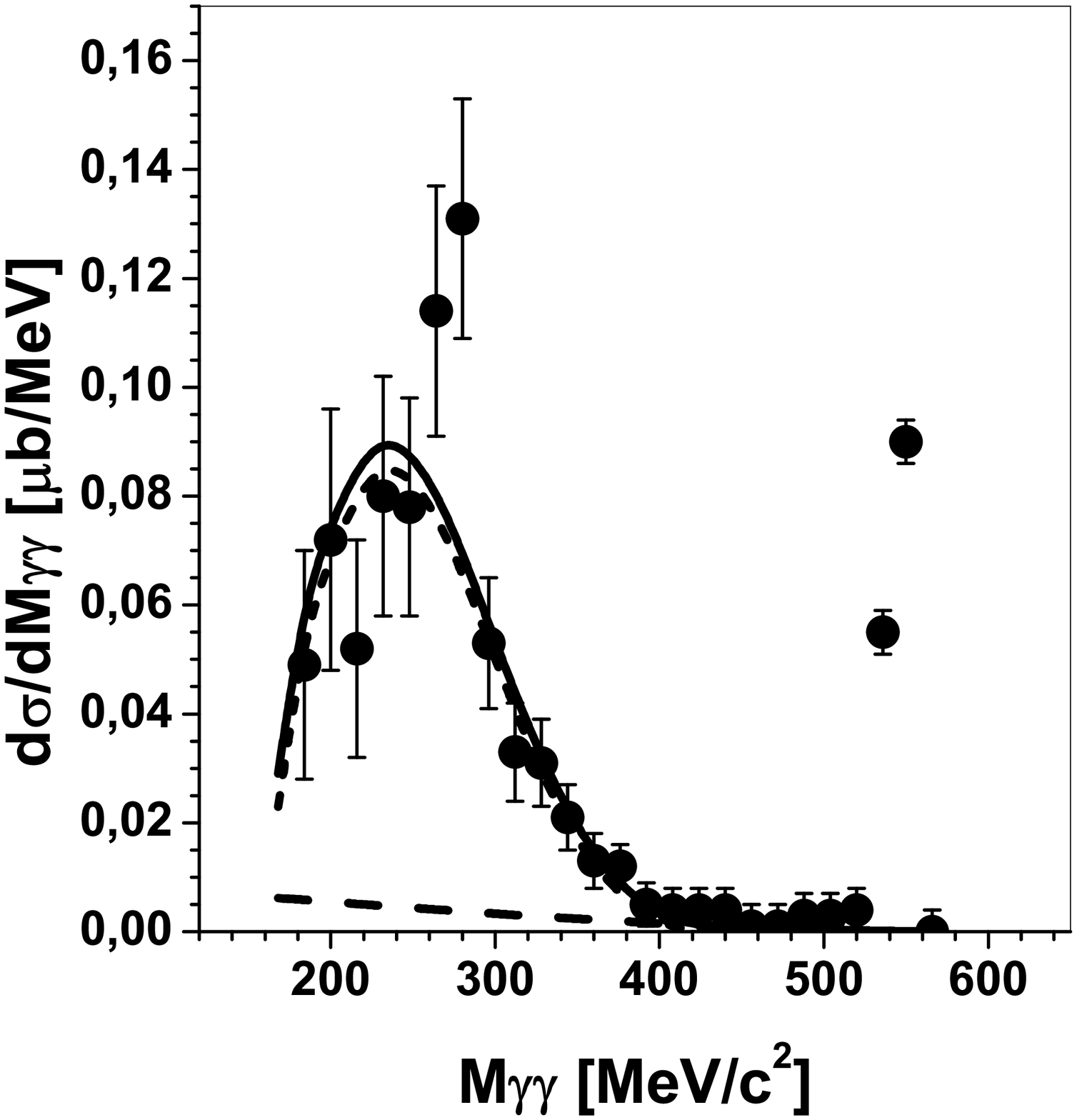}
\end{minipage}
\hspace{\fill}
\begin{minipage}[c]{80mm}
 \caption{\small{Experimentally observed two-photon
invariant mass spectrum of the reaction $pp\gamma\gamma$ (full
circles)\cite{CELWAS} and calculated contributions to this spectrum at 1.36
GeV as a function of $\gamma\gamma$ mass $M_{\gamma \gamma}$. The short dashed line is the contribution from the dibaryon
mechanism. The dashed line is the contribution from the
$\gamma\gamma$ background. The solid line is the
sum of the contributions from the dibaryon mechanism and the
background.}}
\end{minipage}
\vspace*{-0.4cm}
\end{ltxfigure}

 The calculated spectrum was normalized in such a way that the ratio of the value of the calculated total cross section $\sigma_{ee}^{tot}(s)$ for the process (1) at the CMS energy $\sqrt{s}$ of the colliding nucleons
to the total cross section $\sigma_{\gamma}^{tot}(s)$ for the process $NN\to  \gamma d^\star_1$ is equal to the conversion coefficient $R$  given by \cite{Lautrup}
\begin{equation}
R=\frac{\sigma_{ee}^{tot}(s)}{\sigma_{\gamma}^{tot}(s)}=\frac{\alpha}{\pi}\left[\frac{2}{3}\ln\left(\frac{\sqrt{s}- M_R} {m_e}\right)-\frac{2}{9}+O\left(\left(\frac{m_e}{\sqrt{s}- M_R}\right)^2\right)\right].
\end{equation}
Using Eq. (4), for the total cross section $\sigma_{\gamma}^{tot}(s)$ we obtain
\begin{equation}
\sigma_{\gamma}^{tot}(s)=\frac{\omega}{64\pi s p}\frac{1}{4}\int |{\cal{M}}(pp \to \gamma d^\star_1)|^2d\Omega = \frac{C\omega^3} {64\pi s p}[E^2-p^2/3],
\end{equation}
where $\omega=(s-M_R^2)/2\sqrt{s}$, $p$ is the CMS momentum of the colliding protons, $E=\sqrt{p^2+m_p^2}$ and $m_p$ is the proton mass.  The constant $C$ entering into
Eq. (9) was found from the equality $\sigma_{\gamma}^{tot}(s)=\sigma_{\gamma}^{tot}(s_0)$ where $s_0$ is the CMS energy of the $pp$ system for the kinetic energy of incident proton $T_k=1.36$. GeV.
The value $\sigma_{\gamma}^{tot}(s_0)$ was derived from data on the measurement of the two-photon ($\gamma\gamma$) invariant mass distribution of the  $pp \to pp \gamma \gamma$
reaction at $T_k=1.36$ GeV made by the \emph{CELSIUS-WASA Collaboration} \cite{CELWAS}. The surprising feature of this distribution
is a pronounced resonant structure at a mass of
about 280 MeV. It was shown in our work
\cite{Menu07} that this structure is due to the dibaryon mechanism $pp \to \gamma d^\star_1 \to pp \gamma\gamma$ of the studied reaction.
The contribution of this mechanism to the $\gamma\gamma$ invariant mass distribution of the reaction  $pp \to pp \gamma \gamma$ together with the experimentally measured distribution are shown
in Fig.2. Here we also show the $\gamma\gamma$ background calculated under the assumption
that its differential cross section $d\sigma_{back}/dM_{\gamma\gamma}$ is inversely proportional to the $\gamma\gamma$ invariant mass $M_{\gamma\gamma}$. A fit of the sum of the contributions from the dibaryon mechanism and the background to the experimental $\gamma\gamma$ invariant mass distribution gives
for the cross section $\sigma_{\gamma}^{tot}(s_0) = 11\mu$b.
 The total cross section of the background in the mass region of the observed structure was estimated to be 1 $\mu$b.

The dielectron invariant mass spectra  predicted by theoretical models that take into account only the conventional  processes
of dielectron production (solid lines) for the case of  $p+p$ \cite{FFKMppeeX} (left panel) and $C+C$ \cite{BRCas98} (right panel) collisions at 1.04 GeV compared to the corresponding \textit{DLS} data (solid circles) \cite{DLSpppdex} and \cite{DLSAAex}, respectively are presented in Fig. 2.
The short dashed lines show  the dibaryon mechanism contributions calculated by Eq. 6 that includes the form factor (7), and dotted lines show such contributions calculated by Eq. 6 without it. The dashed and dot-dashed lines represent the sums of
the corresponding spectra predicted by the theoretical models and the dibaryon mechanism contributions calculated with and without the form factor (7), respectively.

 Fig. 2 shows that adding the dibaryon mechanism contributions to the  dielectron invariant mass spectra predicted by the theoretical models that take into account only the conventional  processes
of dielectron production we get a very good agreement between resulting spectra and experimental data. Note that the reliability of the numerical values of the calculated contributions of the dibaryon mechanism of dielectron production $NN \to e^+e^- d^*_1$ is provided for by the fact that its total cross section was normalized to the experimentally measured total cross section of the process $pp \to \gamma d^*_1$.
Similar results were obtained for the case of $Ca+Ca$ collisions at 1.04 GeV/c.

The presence or absence of the dibaryon mechanism of $e^+e^-$ pair production can be established by measurement of the missing mass spectrum in the reaction
$pp \to \gamma^\star X \to e^+e^- X$. If this mechanism really exists it should manifest itself in such a spectrum as a narrow peak at a mass around 1956 MeV  superimposed on a smooth background.
\begin{ltxfigure}
\begin{minipage}[c]{80mm}
  \includegraphics[height=.35\textheight]{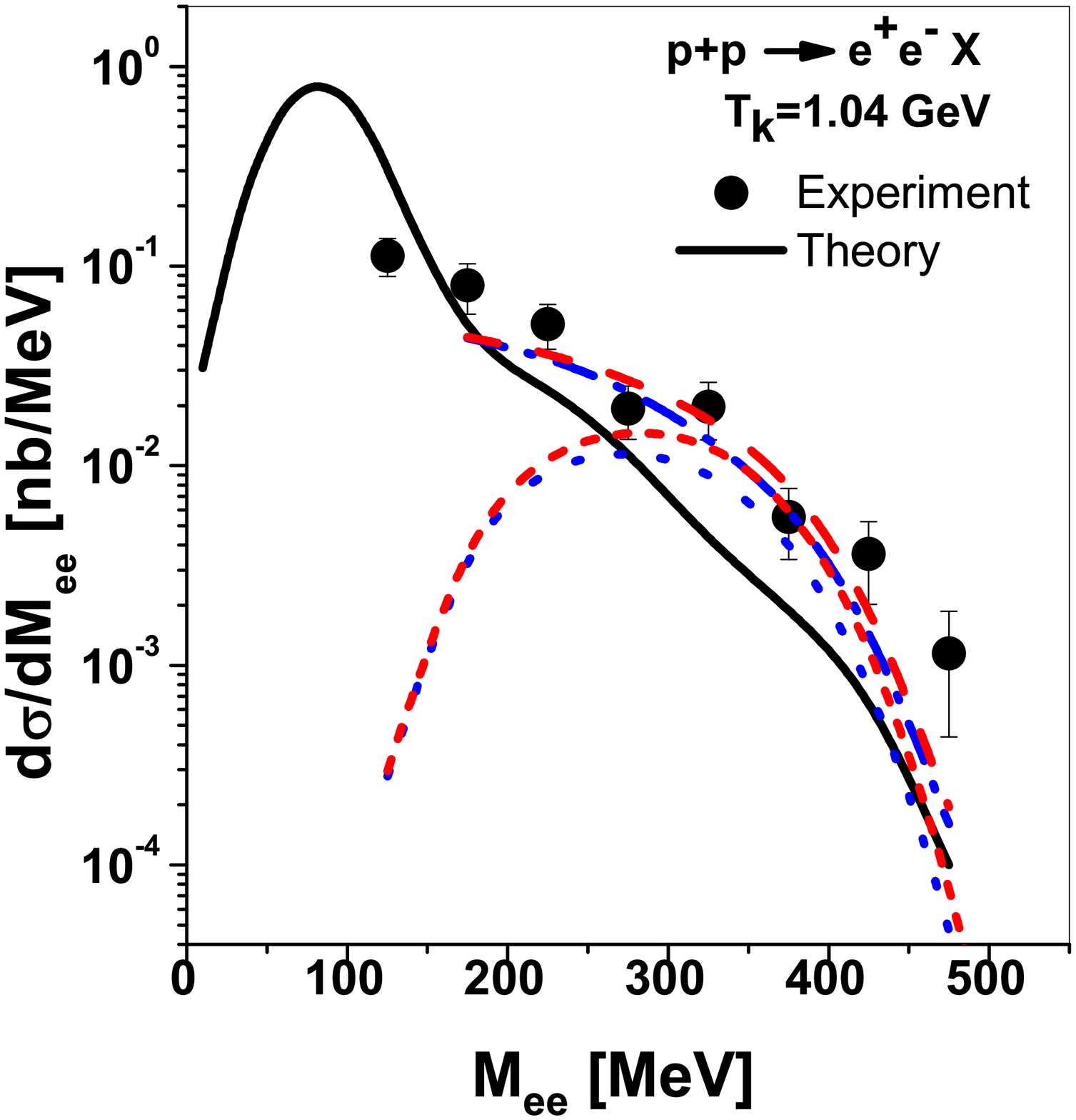}
\end{minipage}
\hspace{\fill}
\begin{minipage}[c]{80mm}
  \includegraphics[height=.35\textheight]{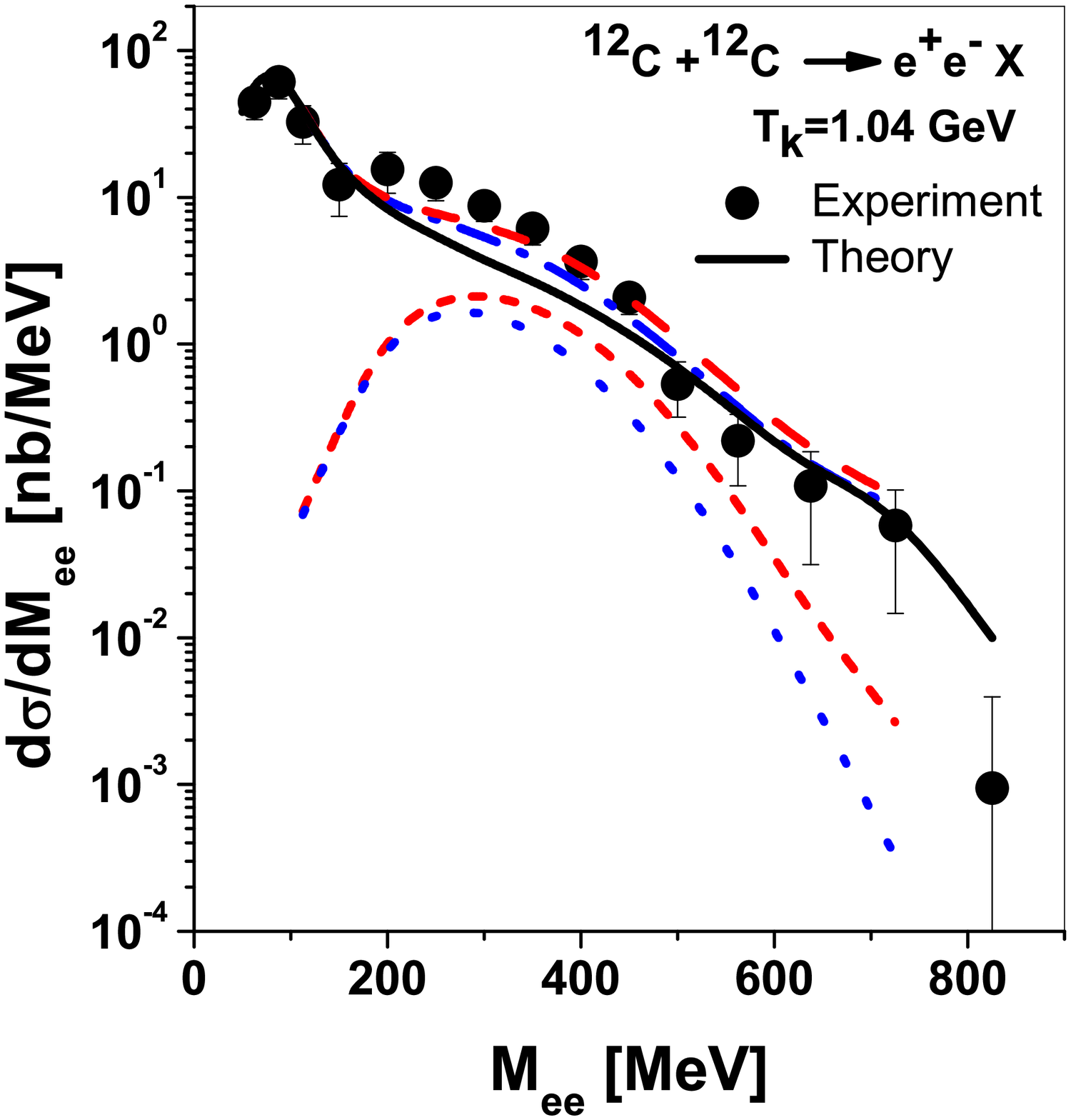}
\end{minipage}
 \caption{\small{The dielectron spectra for $p+p$ (left panel)\cite{DLSpppdex} and $C+C$ (right panel) \cite{DLSAAex} collisions at 1.04 GeV, as a function of the dielectron mass $M_{ee}$ measured by the {\it DLS} compared to
those predicted by the theoretical models that take into account only the conventional  processes
of dielectron production (solid lines) for the $p+p$ \cite{FFKMppeeX} and for $C+C$ \cite{BRCas98} collisions, respectively. The short dashed and dotted lines correspond to the contributions of the dibaryon mechanism calculated by Eq. 6 that includes the form factor (7) and does not include it, respectively. The dashed and dot-dashed lines show the sums of the corresponding spectra predicted by  theoretical models and the dibaryon mechanism contributions with the form factor (7) and without this form factor.}}
\vspace*{-0.4cm}
\end{ltxfigure}
\bibliographystyle{aipproc}   



\end{document}